\documentclass[letterpaper]{article} % DO NOT CHANGE THIS
\usepackage{aaai25}  % DO NOT CHANGE THIS
\usepackage{times}  % DO NOT CHANGE THIS
\usepackage{helvet}  % DO NOT CHANGE THIS
\usepackage{courier}  % DO NOT CHANGE THIS
\usepackage[hyphens]{url}  % DO NOT CHANGE THIS
\usepackage{graphicx} % DO NOT CHANGE THIS
\usepackage{natbib}  % DO NOT CHANGE THIS AND DO NOT ADD ANY OPTIONS TO IT
\usepackage{caption} % DO NOT CHANGE THIS AND DO NOT ADD ANY OPTIONS TO IT
\usepackage{algorithm}
\usepackage{algorithmic}

\usepackage{newfloat}
\usepackage{listings}
\DeclareCaptionStyle{ruled}{labelfont=normalfont,labelsep=colon,strut=off} % DO NOT CHANGE THIS
\floatstyle{ruled}
\newfloat{listing}{tb}{lst}{}
\floatname{listing}{Listing}
\title{EigenSR: Eigenimage-Bridged Pre-Trained RGB Learners for \\ Single Hyperspectral Image Super-Resolution}
\author{
    Xi Su, 
    Xiangfei Shen, 
    Mingyang Wan, 
    Jing Nie, 
    Lihui Chen,\\ 
    Haijun Liu\thanks{Corresponding author. }, 
    Xichuan Zhou\\
}
\affiliations{
    Chongqing University\\
    xisu@stu.cqu.edu.cn, haijun\_liu@126.com
}

\usepackage{bibentry}
\usepackage{subfig}
\usepackage{amsfonts}
\usepackage{algorithmic}
\usepackage{algorithm}
\usepackage{bm}
\usepackage{booktabs}
\usepackage{multirow}
\usepackage{makecell}
\usepackage{diagbox}
\usepackage{graphicx}
\usepackage{pifont}
\usepackage{amsmath}
\usepackage{xcolor}
\usepackage[hang, flushmargin]{footmisc}
\usepackage{colortbl}

\begin{document}

\maketitle

\begin{abstract}
Single hyperspectral image super-resolution (single-HSI-SR) aims to improve the resolution of a single input low-resolution HSI. Due to the bottleneck of data scarcity, the development of single-HSI-SR lags far behind that of RGB natural images. In recent years, research on RGB SR has shown that models pre-trained on large-scale benchmark datasets can greatly improve performance on unseen data, which may stand as a remedy for HSI. But how can we transfer the pre-trained RGB model to HSI, to overcome the data-scarcity bottleneck? Because of the significant difference in the channels between the pre-trained RGB model and the HSI, the model cannot focus on the correlation along the spectral dimension, thus limiting its ability to utilize on HSI. Inspired by the HSI spatial-spectral decoupling, we propose a new framework that first fine-tunes the pre-trained model with the spatial components (known as eigenimages), and then infers on unseen HSI using an iterative spectral regularization (ISR) to maintain the spectral correlation. The advantages of our method lie in: 1) we effectively inject the spatial texture processing capabilities of the pre-trained RGB model into HSI while keeping spectral fidelity, 2) learning in the spectral-decorrelated domain can improve the generalizability to spectral-agnostic data, and 3) our inference in the eigenimage domain naturally exploits the spectral low-rank property of HSI, thereby reducing the complexity. This work bridges the gap between pre-trained RGB models and HSI via eigenimages, addressing the issue of limited HSI training data, hence the name EigenSR. Extensive experiments show that EigenSR outperforms the state-of-the-art (SOTA) methods in both spatial and spectral metrics.
\end{abstract}

\begin{links}
    \link{Code}{https://github.com/enter-i-username/EigenSR}
    % \link{Datasets}{https://aaai.org/example/datasets}
    % \link{Extended version}{https://aaai.org/example/extended-version}
\end{links}

\begin{figure}[t]
  \centering
  \includegraphics[width=\columnwidth]{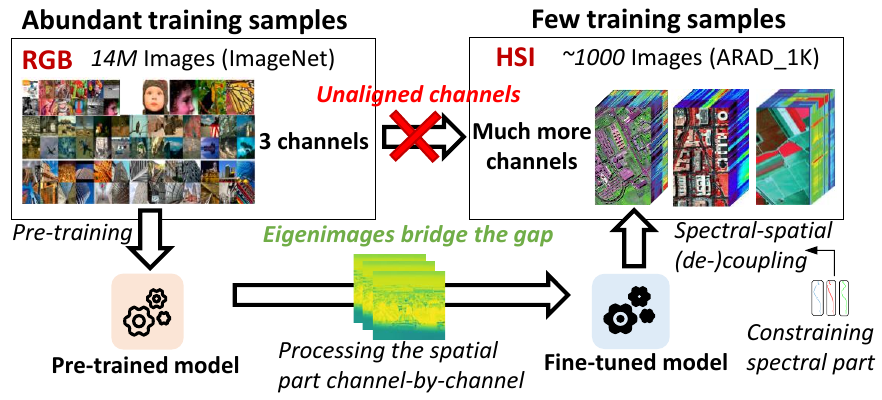}
  \caption{The motivation. The lack of HSI training data is the main bottleneck for single-HSI-SR. The channel differences between RGB and HSI make it challenging to directly transfer RGB models to HSI. Inspired by spectral-spatial decoupling, we leverage spatial part of HSI, namely eigenimages, to bridge the gap between pre-trained RGB models and HSI, and constrain the spectral part to keep spectral fidelity. } 
  \label{fig:motivation}
\end{figure}

\section{Introduction}

Hyperspectral imaging sensors acquire three-dimensional hyperspectral images (HSIs) by capturing a vector with hundreds of contiguous spectral elements from each pixel within a given scene,  offering the perception capacity of various types of real-world materials \cite{Review_HU_Keshava, Review_HU_Bioucas_Challenge}. However, physical restrictions in imaging sensors lead to the final product of low spatial resolution HSI \cite{Review_Hy_Hong, Review_SR_Yokoya}. To tackle this restriction, hyperspectral image super-resolution (HSI-SR) task \cite{Review_SR_Vivone} has received extensive attention by injecting abundant spatial details to HSI for spatial resolution improvement. Typically, HSI-SR can be categorized into two branches of methods: fusion-based \cite{SR_TenSR_Xu_TGRS, SR_NLLR_HyperLap_Peng, SR_TV_TD_Xu} and single-image-based methods \cite{SR_Blind_RefImg_Wan, SR_Blind_SwinIR_SHISR, SR_Blind_Model_Dong, ESSAformer}. In this work, we focus on the single-HSI-SR task.

Similar to the definition of single RGB image SR \cite{RGB_SR_CNN1, SR_Blind_RCAN}, single-HSI-SR aims to recover spatial details with only a single low-resolution (LR) image. With the booming of deep learning \cite{bai1, bai2, bai3, nie1, nie2, nie3}, and the development of large-scale benchmark datasets \cite{ImageNet, DIV2K}, single RGB image SR has reached a level of maturity, exemplified in the studies that leverage pre-training techniques \cite{IPT, EDT, HAT}. Unfortunately, such an effective strategy cannot be applied to HSI due to the scarcity of data \cite{ref_data_starved_1, ref_data_starved_2}. Most existing single-HSI-SR methods can only be trained on a small number of indoor and outdoor images, or on low-resolution sub-images of remote sensing data \cite{SSPSR, ESSAformer}, thus restricting their capabilities. This scarcity of training data has become the major bottleneck in the field of single-HSI-SR. 

Given the lack of a large-scale dataset for pre-training an HSI model, a natural question arises: \textbf{\textit{how can we leverage pre-trained RGB models to feed the data-hungry bottleneck of HSI?}} Directly transferring the RGB model for HSI-SR is difficult, because the channels are not aligned (3 for RGB and much more for HSI) \cite{rgb_transfer_hsi, SR_Blind_SwinIR_SHISR}. Thus the RGB model cannot handle the correlation along the spectral dimension. 

To address the issue, we borrow the idea of HSI spatial-spectral decoupling using the Singular Value Decomposition (SVD) \cite{FastHyMix}, and operate SR using pre-trained RGB models solely on the spatial components (known as eigenimages \cite{fastHyDe}), while constraining the spectral correlations with the spectral components (called the spectral basis), as shown in Figure \ref{fig:motivation}. Specifically, our method consists of two stages. First, we fine-tune the pre-trained RGB model to adapt to the eigenimage domain in a single-channel manner. Second, when the fine-tuned model is used to infer\footnote{In this paper, we do not particularly differentiate between ``test'' and ``infer'', both of which refer to performing single-image super-resolution on the unseen low-resolution HSI.} on the unseen HSI, we decompose the LR HSI, perform SR on the eigenimages channel-by-channel, and reconstruct the SR HSI with the spectral basis. Additionally, an iterative spectral regularization (ISR) is proposed to enhance the spectral correlation by fully utilizing the LR HSI. The advantages lie in threefold: 1) we effectively inject the spatial texture processing capabilities learned from abundant RGB images into HSI while maintaining spectral correlation, 2) due to the learning in the spectral-decorrelated domain, the generalizability to spectral-agnostic data can be improved, and 3) our inference in the eigenimage domain naturally exploits the spectral low-rank property of HSI, thereby reducing the complexity. This work bridges the gap between pre-trained RGB models and HSI through eigenimages, addressing the issue of limited HSI training data, hence the name EigenSR. Extensive experiments show that EigenSR outperforms the SOTA methods in both spatial and spectral metrics.

Our main contributions can be summarized as follows:
\begin{itemize}
    \item We propose a new framework for single-HSI-SR that first introduces large-scale pre-trained RGB models into HSI. It leverages pre-trained models learned from abundant RGB images to address the data-scarcity problem of HSI.

    \item Our method is based on the HSI spatial-spectral decoupling. We inject the spatial texture knowledge provided by the pre-trained model to the spatial part via fine-tuning, and regularize the spectral correlation when inferring on unseen data. This effectively leverages the pre-trained model while maintaining the spectral fidelity.

    \item Extensive experiments demonstrate that our method surpasses the current SOTA methods in both spatial and spectral metrics, validating the effectiveness of introducing the pre-trained RGB models.
\end{itemize}

\section{Related Work}

\subsection{Single RGB Image Super-Resolution}

Single RGB image SR has been a long-standing topic of interest. Compared to convolutional neural networks (CNNs) for SR \cite{RGB_SR_CNN1, RGB_SR_CNN2, RGB_SR_CNN3}, Transformer-based architectures \cite{SR_DL_SwinIR_Liang, RGB_SR_ViT2, HAT, RGB_SR_ViT4} have recently gained prominence in SR due to their ability to extract non-local spatial features and their scalability. Recent studies have found that self-supervised pre-training on large benchmark datasets, such as ImageNet \cite{ImageNet}, followed by fine-tuning on high-quality domain-specific datasets, such as DIV2K \cite{DIV2K} and Flickr2K \cite{Flickr2K}, significantly enhances the Transformer's power to process spatial information \cite{IPT, EDT, HAT}. However, HSI lacks such large datasets, which hinders the application of pre-training. Therefore, we introduce pre-trained RGB models for HSI to address this data bottleneck.

\subsection{Single HSI Super-Resolution}

Single-HSI-SR not only involves processing spatial information but also focuses on the spectral correlation. Researchers used the 3D-CNN \cite{2D3DHSISR}, group convolution \cite{SSPSR}, spectrum and feature context \cite{SFCSR}, recurrent mechanism \cite{RFSR}, and Transformers \cite{MSDformer, ESSAformer} to learn the spatial-spectral correlations in HSI in an end-to-end manner. Limited by the data, these models may not reach their full potential. 

To address this issue, two interesting attempts \cite{rgb_transfer_hsi, SR_Blind_SwinIR_SHISR} have found the effectiveness of leveraging RGB images for HSI-SR, inspired by transfer learning. They trained single-channel models on RGB images and inferred on HSI band-by-band to align the channels, followed by a non-negative matrix factorization (NMF) to ensure the spectral fidelity. However, the band-by-band operation in original HSI space and optimization of NMF may result in high time costs. Unlike them, we perform SR in the eigenimage domain, which more effectively and efficiently leverages large-scale pre-trained models.

\subsection{Learning in the Eigenimage Domain}

Learning in the eigenimage domain enables the independent processing of spectral or spatial components. Current denoising methods leverage the natural low-rank property of eigenimages and have achieved impressive results. For instance, off-the-shelf methods \cite{BM3D} were inserted into eigenimages for efficient denoising and inpainting \cite{fastHyDe}. CNNs were used on eigenimages to remove mixed noise in HSI \cite{FastHyMix}. Deep spatial priors were learned exclusively in the eigenimage domain for HSI denoising \cite{coefficent_denoiser}. Inspired by the spectral-spatial decoupling, we perform single-HSI-SR on eigenimages in this work. We will demonstrate the effectiveness of eigenimages in the field of SR through theoretical analysis and experimental observations.

\section{Methodology}
\subsection{Problem Formulation}

Single-HSI-SR learns a SR model $f_{SR}(\cdot)$ on the training data, mapping the low-resolution HSI $\mathbf{Y}_{LR} \in \mathbb{R}^{L \times (N/\kappa^{2})}$ to the high-resolution (HR) HSI $\mathbf{Y}_{HR} \in \mathbb{R}^{L \times N}$, where $L$, $N$, and $\kappa$ represent the number of spectral bands, the number of pixels, and the spatial downsampling rate, respectively. Since there is no real dataset of LR-HR pairs, we follow the common practice that uses a downsampling operation to generate LR HSI from HR HSI by: $\mathbf{Y}_{LR} = \mathbf{Y}_{HR} \mathbf{D}$, where $\mathbf{D} \in \mathbb{R}^{N \times (N/\kappa^{2})}$ is the linear downsampling operator. Specifically, the downsampling operator $\mathbf{D}$ is the Bicubic function \cite{HAT, RGB_SR_ViT4}.

Limited by the data scarcity, $f_{SR}(\cdot)$ cannot generalizes well on unseen data. To address this limitation, we introduce the pre-trained RGB model into HSI. However, due to the difference in the image channels, RGB models cannot capture the spectral correlations in HSI. Next, we will explain how to overcome this using eigenimages.

\subsubsection{SR with Eigenimages.} Singular Value Decomposition (SVD) decomposes an HSI $\mathbf{Y}$ into spectral and spatial components. This can be described as:
\begin{equation}
\{\mathbf{U}, \bm{\Sigma}, \mathbf{V}^{\top}\} = \text{SVD}(\mathbf{Y}), \text{or } \mathbf{Y} = \mathbf{U} \bm{\Sigma} \mathbf{V}^{\top}, 
\end{equation}
where $(\cdot)^{\top}$ is the transpose operation; $\mathbf{U} \in \mathbb{R}^{L \times L}$ and $\mathbf{V} \in \mathbb{R}^{N \times L}$ are semiunitary matrices, satisfying $\mathbf{U}^{\top} \mathbf{U} = \mathbf{I}$ and $\mathbf{V}^{\top} \mathbf{V} = \mathbf{I}$; $\bm{\Sigma} = \text{diag}\{\sigma_1, \dots, \sigma_L\}$ is a diagonal matrix of singular values arranged in descending order. We combine $\bm{\Sigma}$ and $\mathbf{V}^{\top}$ to form $\mathbf{E} = \bm{\Sigma} \mathbf{V}^{\top}$, and the rows $\mathbf{E}_{i, :}$ in $\mathbf{E}$ are referred to as a set of \textit{eigenimages} \cite{fastHyDe}. In this way, $\mathbf{Y}$ can be decomposed into orthogonal basis vectors $\mathbf{U}_{:, i}$ representing spectral information and corresponding eigenimages $\mathbf{E}_{i, :}$ representing spatial information. Due to the semiunitary property of $\mathbf{U}$, eigenimages can also be represented as the projection form $\mathbf{E}_{i, :} = (\mathbf{U}_{:, i})^{\top} \mathbf{Y}$.

We rewrite the SVD of $\mathbf{Y}$ in the following form:
\begin{equation}
\label{eq:lowrank_approx}
\mathbf{Y} = \sum_{i=1}^{L} \sigma_{i} \mathbf{U}_{:, i} (\mathbf{V}^{\top})_{i, :} = \sum_{i=1}^{L} \mathbf{U}_{:, i} \mathbf{E}_{i, :} \approx \sum_{i=1}^{p} \mathbf{U}_{:, i} \mathbf{E}_{i, :},
\end{equation}
where we have $p \ll L$. The rank-$p$ approximation (\ref{eq:lowrank_approx}) holds for most real-world HSIs \cite{lowrank}, as they may live in a low-dimensional space. Thus, $\mathbf{Y}$ can be well reconstructed by: $\mathbf{Y} \approx \mathbf{U}_{:, 1:p} ((\mathbf{U}_{:, 1:p})^{\top} \mathbf{Y}$).

Now, through the above analysis, we derive an efficient method for manipulating the HSI in the eigenimage domain rather than in the original space:
\begin{equation}
\label{eq:useeigen}
\left\{
\begin{aligned}
    \mathbf{U} &= \text{SVD}(\mathbf{Y}), \\
    \hat{\mathbf{Y}} &= \mathbf{U}_{:, 1:p} f_{SR}((\mathbf{U}_{:, 1:p})^{\top} \mathbf{Y}),
\end{aligned}
\right.
\end{equation}
where we use only the $\mathbf{U}$ matrix from the SVD for simplicity; $f_{SR}(\cdot)$ is the SR model that super-resolves the eigenimages. Specifically, we perform SR with the eigenimages channel-by-channel, to align the channel difference between the RGB model and eigenimages. 

The advantages of SR with eigenimages are: 1) thanks to spectral decoupling, the SR model can focus solely on spatial information, while spectral correlations are stored in $\mathbf{U}_{:, 1:p}$ to restore the original HSI space. This ensures that the super-resolved HSI maintains spectral fidelity; 2) eigenimages only contain the spatial information, which is more generalizable to spectral-agnostic unseen data; 3) processing $p$ channels is more efficient than $L$ in the original HSI space.

\begin{figure}[!t]
  \centering
  % scale=0.55
  % width=\columnwidth 
  \includegraphics[scale=0.54]{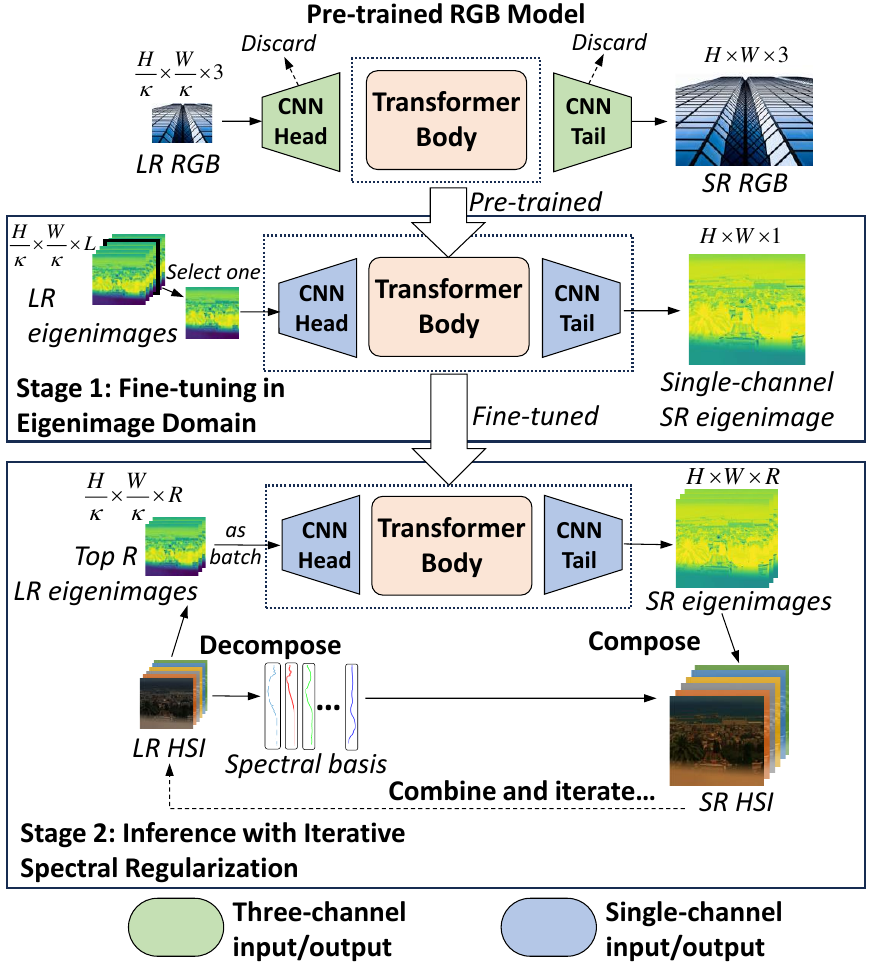}
    \caption{The flowchart of EigenSR. It consists of two stages. Stage 1 uses the pre-trained Transformer Body to fine-tune a single-channel model in the eigenimage domain. Stage 2 utilizes the fine-tuned model for inference with iterative spectral regularization on unseen LR HSI.}
  \label{fig:flowchart}
\end{figure}

\subsection{The Proposed EigenSR}

Now, we will explain in detail how our EigenSR fully leverages the pre-trained RGB model to process HSI based on eigenimages. Our method consists of two stages, as shown in Figure \ref{fig:flowchart}. First, we fine-tune the pre-trained model using eigenimages. Second, when the model is applied to unseen data, we use Eq. (\ref{eq:useeigen}) to perform SR and introduce an iterative spectral regularization (ISR). By repeatedly performing model inference, updating $\mathbf{U}$, and taking the weighted average of the SR results, we achieve better spectral fidelity.

\subsubsection{Fine-Tuning the Pre-Trained Model in the Eigenimage Domain.} Transformer-based architectures generally consist of a Transformer body and a CNN Head and Tail, with the Head and Tail serving as three-channel input and output layers specifically for RGB images. There are two reasons for fine-tuning: 1) to replace the input and output channels of the Head and Tail with one channel to align with the channels of eigenimages and focus solely on processing spatial details, and 2) to adapt to the data in the eigenimage domain. We list the key steps of the fine-tuning process below:

\noindent\textit{\textbf{Step 1:}} Prepare in advance the training set of triplets $\{(\mathbf{Y}_{HR}, \mathbf{Y}_{LR}, \mathbf{U}_{HR})_{n}\}_{n}$, where each $\mathbf{Y}_{HR}$ is the available HR HSI, $\mathbf{Y}_{LR} = \mathbf{Y}_{HR} \mathbf{D}$ is the generated LR HSI, and $\mathbf{U}_{HR} = \text{SVD}(\mathbf{Y}_{HR})$ is the spectral basis from $\mathbf{Y}_{HR}$;

\noindent\textit{\textbf{Step 2:}} In each epoch, randomly select one channel $c$, and obtain $(\mathbf{E}_{HR})_{c, :} = ((\mathbf{U}_{HR})_{:, c})^{\top} \mathbf{Y}_{HR}$ and $(\mathbf{E}_{LR})_{c, :} = ((\mathbf{U}_{HR})_{:, c})^{\top} \mathbf{Y}_{LR}$, by projecting both $\mathbf{Y}_{HR}$ and $\mathbf{Y}_{LR}$ onto the same eigenimage space using $((\mathbf{U}_{HR})_{:, c})^{\top}$.

\noindent\textit{\textbf{Step 3:}} Feed the model with the LR eigenimage $(\mathbf{E}_{LR})_{c, :}$ and yield the super-resolved eigenimage $(\mathbf{E}_{SR})_{c, :}$, by $(\mathbf{E}_{SR})_{c, :} = f_{SR}((\mathbf{E}_{LR})_{c, :})$.

\noindent\textit{\textbf{Step 4:}} Calculate the loss between $(\mathbf{E}_{SR})_{c, :}$ and $(\mathbf{E}_{HR})_{c, :}$, and perform backward propagation.

In this work, we use the pre-trained Image Processing Transformer (IPT) \cite{IPT} as the pre-trained RGB model. The loss function is the L1 loss. For the settings of fine-tuning, please refer to the Experiments section.

\subsubsection{Why Eigenimages Work?} Here we illustrate the reason behind the effectiveness of learning eigenimages with the following theorem.

\noindent\textbf{Theorem 1}: For eigenimages $\mathbf{E} = \mathbf{U}^{\top} \mathbf{Y}$, we have:	
\begin{enumerate}
\item If two column vectors $\mathbf{Y}_{:, i}$ and $\mathbf{Y}_{:, j}$ of $\mathbf{Y}$ are identical, then their corresponding $\mathbf{E}_{:, i}$ and $\mathbf{E}_{:, j}$ are also identical.
\item If $\mathbf{Y}$ is downsampled by a linear operator $\mathbf{D}$, then the same $\mathbf{D}$ concurrently acts on $\mathbf{E}$.
\end{enumerate}	 

Theorem 1 indicates that each channel $\mathbf{E}_{i, :}$ of $\mathbf{E}$ preserves the spatial structure of the original space $\mathbf{Y}$, and that $\mathbf{D}$ acts on each $\mathbf{E}_{i, :}$. This enables us to fine-tune a single-channel model for $\mathbf{E}$, which has already been pre-trained on RGB datasets, to learn the inverse mapping of the same degradation $\mathbf{D}$. The proof of Theorem 1 is given in the Appendix. 

\begin{figure}[!t]
  \centering
  % \hfill
  \subfloat[]{\includegraphics[scale=0.33]{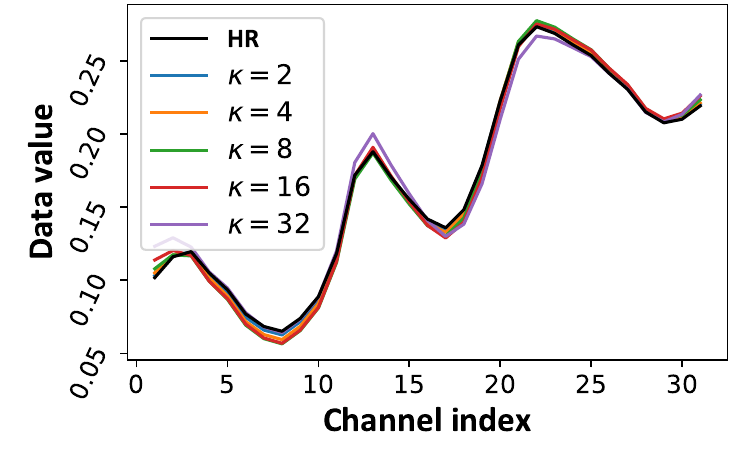}%
  \label{fig_beads_u1}}
  \hfil
  \subfloat[]{\includegraphics[scale=0.33]{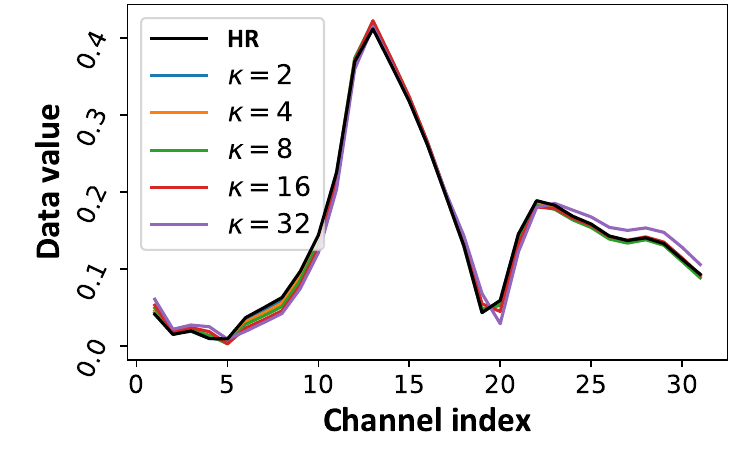}%
  \label{fig_beads_u2}}
  \caption{Absolute spectral basis of the CAVE-beads image at different downsampling rates. (a) $| \mathbf{U}_{:, 1} |$ and (b) $| \mathbf{U}_{:, 2} |$.}
  \label{fig:spectralreason}
\end{figure}

\begin{algorithm}[t]
\caption{Inference with iterative spectral regularization.}
\label{alg:inference}
\begin{algorithmic}[1]
\renewcommand{\algorithmicrequire}{\textbf{Input:}}
\REQUIRE

Observed low-resolution HSI $\mathbf{Y}_{LR}$; SR model $f_{SR}(\cdot)$; Rank $R$; Iteration number $N_{it}$; Combination weight constant $\lambda$.%Geometric transform sequence $\{\mathcal{G}^{(1)}, \ldots, \mathcal{G}^{(N_{it})}\}$ and their inverses $\{\mathcal{G}^{-1 (1)}, \ldots, \mathcal{G}^{-1 (N_{it})}\}$; 

\renewcommand{\algorithmicensure}{\textbf{Output:}}
\ENSURE
 The predicted SR HSI $\mathbf{Y}_{comb}^{(N_{it}+1)}$.
 
\STATE $\mathbf{Y}_{comb}^{(1)} = \mathbf{Y}_{LR}$;

\FOR{$i$ \textbf{in} $\{1, \ldots, N_{it}\}$}

\STATE $\mathbf{U}^{(i)} = \text{SVD}(\mathbf{Y}_{comb}^{(i)})$;

\STATE $\mathbf{E}_{LR}^{(i)} = (\mathbf{U}_{:, 1:R}^{(i)})^{\top} \mathbf{Y}_{LR}$;

% \STATE $\mathbf{E}_{LR}^{(i)} = \mathcal{G}^{(i)}(\mathbf{E}_{LR}^{(i)})$;
\textcolor{gray}{\# The channel-by-channel processing in lines 5-7 can be executed in parallel within a batch. }
\FOR{$j$ \textbf{in} $\{1, \ldots, R\}$}

\STATE $(\mathbf{E}_{SR}^{(i)})_{j,:} = f_{SR}((\mathbf{E}_{LR}^{(i)})_{j,:})$; 

\ENDFOR

% \STATE $\mathbf{E}_{SR}^{(i)} = \mathcal{G}^{-1 (i)}(\mathbf{E}_{SR}^{(i)})$;

\STATE $\mathbf{Y}_{SR}^{(i)} = \mathbf{U}_{:, 1:R}^{(i)} \mathbf{E}_{SR}^{(i)}$;

\STATE $\mathbf{Y}_{comb}^{(i+1)} = \lambda \mathbf{Y}_{SR}^{(i)} + (1 - \lambda) \mathbf{Y}_{comb}^{(i)}$; \textcolor{gray}{\# We replace $\mathbf{Y}_{comb}^{(1)}$ here with $\text{Bicubic}\uparrow_{\kappa}(\mathbf{Y}_{comb}^{(1)})$.}

\ENDFOR

\end{algorithmic}
\end{algorithm}

\subsubsection{Inference with Iterative Spectral Regularization on Unseen Data.} The high-quality spectral basis $\mathbf{U}_{HR}$ from $\mathbf{Y}_{HR}$ can be used during fine-tuning. However, during inference on unseen data, only the LR HSI is available. We found that the dominant components of $\mathbf{U}$ (e.g., $\mathbf{U}_{:, 1}$ and $\mathbf{U}_{:, 2}$) at different downsampling rates are not affected seriously, as shown in Figure \ref{fig:spectralreason}. This is because most spectral information in $\mathbf{Y}_{LR}$ is preserved despite the spatial degradation, even in the extreme case where $\kappa = 32$. Therefore, by assuming that the spectral information in $\mathbf{Y}_{LR}$ and $\mathbf{Y}_{HR}$ is close \cite{rgb_transfer_hsi}, we can extract the spectral information from  $\mathbf{Y}_{LR}$ to faithfully reconstruct $\mathbf{Y}_{HR}$. We thereby achieve an efficient method \textbf{EigenSR-}$\bm{\alpha}$, through:
\begin{equation}
\left\{
\begin{aligned}
    \mathbf{U} &= \text{SVD}(\mathbf{Y}_{LR}), \\
    \mathbf{Y}_{SR} &= \mathbf{U}_{:, 1:R} f_{SR}((\mathbf{U}_{:, 1:R})^{\top} \mathbf{Y}_{LR}),
\end{aligned}
\right.
\end{equation}

\begin{table}[!b]
\centering
\renewcommand{\arraystretch}{0.6}
\scriptsize
% \small
\tabcolsep=0.03cm
% \resizebox{\columnwidth}{!}{
\begin{tabular}{c|c c c c c}
\toprule
\textbf{Dataset} & \textbf{Type} & \makecell{\textbf{\#test/train}} & \textbf{Sensor} & \textbf{Wavelength} & \textbf{Size} \\
\midrule
\rowcolor{gray!10}
ARAD\_1K & HSI & 50/900 & Specim IQ & 400-700nm & $482 \times 512 \times 31$ \\
CAVE & HSI & 32/- & Apogee Alta U260 & 400-700nm & $512 \times 512 \times 31$ \\
\rowcolor{gray!10}
Harvard & HSI & 50/- & Nuance FX & 420-720nm & $512 \times 512 \times 31$ \\
Pavia & HSI & 5/- & ROSIS & 430-860nm & $192 \times 192 \times 102$ \\
\rowcolor{gray!10}
DC Mall & HSI & 8/- & Hydice & 400-2400nm & $128 \times 128 \times 191$ \\
Chikusei & HSI & 48/- & HH-VNIR-C & 363-1018nm & $128 \times 128 \times 128$ \\
\rowcolor{gray!10}
RESISC45 & RGB & -/29250 & - & Visible light & $256 \times 256 \times 3$ \\
\bottomrule
\end{tabular}
% }
\caption{Dataset information.}
\label{tab:dataset}
\end{table}

\noindent where $\mathbf{Y}_{SR}$ is the SR HSI; $f_{SR}(\cdot)$ is the model fine-tuned on eigenimages that operates channel-by-channel; $R$ is the rank hyperparameter balancing the reconstruction performance and inference efficiency. Although $\mathbf{U}$ assists to maintain the spectral correlations in the original HSI space, it contains compressed and incomplete spectral information of $\mathbf{Y}_{LR}$. We further propose a new iterative spectral regularization (ISR) to fully utilize the spectral correlations of $\mathbf{Y}_{LR}$, by combining the result from the last iteration and then updating a new spectral basis as the following:
\begin{equation}
\left\{
\begin{aligned}
    \mathbf{Y}_{comb}^{(i+1)} &= \lambda \mathbf{Y}_{SR}^{(i)} + (1 - \lambda) \mathbf{Y}_{comb}^{(i)}, \\
    \mathbf{U}^{(i+1)} &= \text{SVD}(\mathbf{Y}_{comb}^{(i+1)}),
\end{aligned}
\right.
\end{equation}
where $\lambda \in (0, 1]$ is a weight constant; $\mathbf{Y}_{comb}^{(i)}$ is the combined result starting from $\mathbf{Y}_{LR}$. The detailed explanation is in Algorithm~\ref{alg:inference}. We denote this version as \textbf{EigenSR-}$\bm{\beta}$. Note that EigenSR-$\beta$ takes more inference steps, but has increased spectral regularization. When we set $\lambda$ and the iteration number $N_{it}$ to 1, EigenSR-$\beta$ degenerates to EigenSR-$\alpha$.

\begin{table*}[!t]
\centering
\renewcommand{\arraystretch}{0.53}
\scriptsize
% \small
\tabcolsep=0.26cm
% \textwidth
% \resizebox{\textwidth}{!}{
\begin{tabular}{l|c|c c c|c c c|c c c}

\toprule
\multirow{2}{*}{\textbf{Method}} & \multirow{2}{*}{\textbf{Scale}} & \multicolumn{3}{c|}{\textbf{ARAD\_1K-Test}} & \multicolumn{3}{c|}{\textbf{CAVE}} & \multicolumn{3}{c}{\textbf{Harvard}} \\
% \cline{3-11}
& & \textbf{PSNR}$\uparrow$ & \textbf{SSIM}$\uparrow$ & \textbf{SAM}$\downarrow$ & \textbf{PSNR}$\uparrow$ & \textbf{SSIM}$\uparrow$ & \textbf{SAM}$\downarrow$ & \textbf{PSNR}$\uparrow$ & \textbf{SSIM}$\uparrow$ & \textbf{SAM}$\downarrow$ \\
\midrule

\rowcolor{gray!10} % 10% gray
Bicubic                        & $\times 2$ &43.76&0.9824&0.878 &40.11&0.9770&2.905 &41.05&0.9545&2.314 \\
DHSP \cite{DHSP}               & $\times 2$ &43.67&0.9804&2.016 &42.40&0.9769&4.025 &42.45&0.9604&2.496 \\
\rowcolor{gray!10}
SSPSR \cite{SSPSR}             & $\times 2$ &48.06&0.9913&0.826 &42.95&0.9842&3.217 &44.58&0.9750&2.331 \\
SFCSR \cite{SFCSR}             & $\times 2$ &48.70&\underline{0.9921}&\underline{0.656} &\underline{44.49}&\underline{0.9865}&\underline{2.612} &\underline{45.56}&\underline{0.9797}&\underline{2.012} \\
\rowcolor{gray!10}
RFSR \cite{RFSR}               & $\times 2$ &48.25&0.9914&0.894 &43.73&0.9855&2.922 &45.20&0.9783&2.086 \\
TSBSR \cite{SR_Blind_SwinIR_SHISR}&$\times 2$&48.12&0.9917&0.658 &44.00&0.9860&2.654 &43.20&0.9707&2.045 \\
\rowcolor{gray!10}
MSDformer \cite{MSDformer}     & $\times 2$ &47.44&0.9908&1.011 &42.53&0.9829&3.457 &44.49&0.9748&2.321 \\
ESSAformer \cite{ESSAformer}   & $\times 2$ &\underline{48.97}&\underline{0.9927}&0.714 &43.56&0.9857&2.909 &44.93&0.9767&2.265 \\
\rowcolor{gray!10}
DKP \cite{DKP_CVPR24}          & $\times 2$ &44.16&0.9827&1.957 &42.96&0.9770&4.070 &42.70&0.9650&2.599 \\
\textbf{EigenSR}-$\bm{\alpha}$ \textbf{(Ours)} & $\times 2$ &\underline{49.09}&\textbf{0.9930}&\underline{0.607} &\underline{44.67}&\underline{0.9869}&\underline{2.609} &\underline{45.65}&\underline{0.9805}&\underline{2.006} \\
\rowcolor{gray!10}
\textbf{EigenSR}-$\bm{\beta}$ \textbf{(Ours)}  & $\times 2$ &\textbf{49.11}&\textbf{0.9930}&\textbf{0.600} &\textbf{44.97}&\textbf{0.9874}&\textbf{2.448} &\textbf{45.77}&\textbf{0.9819}&\textbf{1.924} \\
\midrule

\rowcolor{gray!10}
Bicubic                        & $\times 4$ &37.08&0.9280&1.641 &34.38&0.9312&4.208 &36.12&0.8811&2.741 \\
DHSP \cite{DHSP}               & $\times 4$ &35.75&0.9257&1.707 &36.26&0.9379&6.052 &37.15&0.8911&2.922 \\
\rowcolor{gray!10}
SSPSR \cite{SSPSR}             & $\times 4$ &39.87&0.9524&1.713 &37.11&0.9538&4.505 &39.84&0.9287&2.716 \\
SFCSR \cite{SFCSR}             & $\times 4$ &40.06&0.9535&1.511 &37.65&0.9562&3.989 &39.97&0.9296&2.649 \\
\rowcolor{gray!10}
RFSR \cite{RFSR}               & $\times 4$ &39.91&0.9526&1.823 &37.39&0.9551&4.474 &39.97&0.9299&2.665 \\
TSBSR \cite{SR_Blind_SwinIR_SHISR}&$\times 4$&39.81&0.9548&\underline{1.280} &\underline{37.77}&\underline{0.9583}&\underline{3.715} &38.31&0.9134&\underline{2.593} \\
\rowcolor{gray!10}
MSDformer \cite{MSDformer}     & $\times 4$ &39.00&0.9437&2.134 &36.03&0.9376&6.186 &38.78&0.9135&2.893 \\
ESSAformer \cite{ESSAformer}   & $\times 4$ &\underline{40.40}&\underline{0.9565}&1.384 &37.48&0.9576&4.144 &\underline{40.04}&\underline{0.9309}&2.661 \\
\rowcolor{gray!10}
DKP \cite{DKP_CVPR24}          & $\times 4$ &38.27&0.9343&1.824 &36.48&0.9401&6.077 &37.66&0.8999&3.005 \\
\textbf{EigenSR}-$\bm{\alpha}$ \textbf{(Ours)} & $\times 4$ &\underline{40.28}&\underline{0.9602}&\underline{1.223} &\underline{38.29}&\underline{0.9619}&\underline{3.576} &\underline{40.18}&\underline{0.9313}&\underline{2.543} \\
\rowcolor{gray!10}
\textbf{EigenSR}-$\bm{\beta}$ \textbf{(Ours)}  & $\times 4$ &\textbf{40.46}&\textbf{0.9605}&\textbf{1.182} &\textbf{38.36}&\textbf{0.9624}&\textbf{3.480} &\textbf{40.21}&\textbf{0.9315}&\textbf{2.524} \\
\midrule

\rowcolor{gray!10}
Bicubic                        & $\times 8$ &32.66&0.8453&2.613 &30.07&0.8563&5.756 &32.53&0.7932&3.222 \\
DHSR \cite{DHSP}               & $\times 8$ &33.08&0.8409&5.083 &31.95&0.8700&8.210 &34.91&0.8501&3.343 \\
\rowcolor{gray!10}
SSPSR \cite{SSPSR}             & $\times 8$ &34.60&0.8770&2.646 &32.60&0.8917&6.120 &36.18&0.8616&3.090 \\
SFCSR \cite{SFCSR}             & $\times 8$ &34.59&0.8760&2.484 &32.75&0.8926&5.514 &36.21&0.8613&3.052 \\
\rowcolor{gray!10}
RFSR \cite{RFSR}               & $\times 8$ &34.51&0.8738&3.222 &32.53&0.8894&7.170 &36.17&0.8607&3.192 \\
TSBSR \cite{SR_Blind_SwinIR_SHISR}&$\times 8$&34.55&0.8810&\underline{2.310} &32.95&0.9000&\underline{5.158} &34.43&0.8309&3.087 \\
\rowcolor{gray!10}
MSDformer \cite{MSDformer}     & $\times 8$ &34.26&0.8705&3.018 &31.04&0.8476&8.851 &35.11&0.8415&3.750 \\
ESSAformer \cite{ESSAformer}   & $\times 8$ &\underline{34.71}&\underline{0.8831}&2.363 &\underline{33.00}&\underline{0.9005}&5.670 &\underline{36.27}&\underline{0.8651}&\underline{3.043} \\
\rowcolor{gray!10}
DKP \cite{DKP_CVPR24}          & $\times 8$ &33.69&0.8517&5.136 &32.23&0.8769&8.424 &34.52&0.8479&3.563 \\
\textbf{EigenSR}-$\bm{\alpha}$ \textbf{(Ours)} & $\times 8$ &\underline{34.62}&\underline{0.8814}&\underline{2.140} &\underline{33.08}&\underline{0.9006}&\underline{5.195} &\underline{36.31}&\underline{0.8689}&\underline{3.009} \\
\rowcolor{gray!10}
\textbf{EigenSR}-$\bm{\beta}$ \textbf{(Ours)}  & $\times 8$ &\textbf{34.75}&\textbf{0.8835}&\textbf{2.086} &\textbf{33.13}&\textbf{0.9008}&\textbf{4.945} &\textbf{36.37}&\textbf{0.8696}&\textbf{2.975} \\
\bottomrule

\end{tabular}
% }
\caption{Quantitative comparison with the SOTA methods with reference image. ARAD\_1K-Train is used for training. The best results are marked in \textbf{bold}, and the second and third are \underline{underlined}. }
\label{tab:sota_ref}
\end{table*}

\section{Experiments}

\subsection{Experimental Settings}

\subsubsection{Datasets.} We list the dataset information in Table \ref{tab:dataset}. The ARAD\_1K \cite{arad1k}, CAVE \cite{cave}, and Harvard \cite{harvard} datasets contain high-quality HSIs of indoor and outdoor scenes, each with 31 spectral bands, collected using three different sensors. The ARAD\_1K dataset includes 950 HSIs, with the first 900 used as the training set (denoted as ARAD\_1K-Train) and the remaining 50 used as the test set (denoted as ARAD\_1K-Test). The CAVE and Harvard datasets contain only a few dozen images, and we used them as test sets.

The remote sensing datasets Pavia\footnote{\scriptsize{https://ehu.eus/ccwintco/index.php?title=Hyperspectral\_Remote\_Sensing\_Scenes}}, DC Mall\footnote{\scriptsize{http://lesun.weebly.com/hyperspectral-data-set.html}}, and Chikusei \cite{chikusei} contain more than 100 channels, but have low spatial resolutions. The RESISC45\footnote{\scriptsize{https://tensorflow.google.cn/datasets/catalog/resisc45}} includes HR RGB images of remote sensing scenes. We used the RESISC45 dataset as the training set, and the Pavia, DC Mall, and Chikusei datasets as test sets to conduct blind testing on the three datasets without the HR reference image.

\subsubsection{Metrics.} We used the peak signal-to-noise ratio (PSNR) and structural similarity (SSIM) to evaluate the spatial quality of SR HSI, and the spectral angle mapper (SAM) for the spectral quality. When testing without reference image, a no-reference quality measurement score (NRQMS) \cite{no_ref_metric} was used to jointly assess the spectral-spatial quality.

\subsubsection{Comparative Methods.} Several methods were compared, including three unsupervised methods Bicubic, DHSP \cite{DHSP}, and DKP \cite{DKP_CVPR24}, and six supervised methods SSPSR \cite{SSPSR}, SFCSR \cite{SFCSR}, RFSR \cite{RFSR}, TSBSR \cite{SR_Blind_SwinIR_SHISR}, MSDformer \cite{MSDformer}, and ESSAformer \cite{ESSAformer}. Among them, TSBSR transfers RGB models to HSI and uses NMF as the spectral constraint. Since it only has remote sensing models, we replaced its model with the fine-tuned IPT on original HSIs to adapt to various scenes.

\subsubsection{Implementation Details.} We used two NVIDIA 3090 cards to run the algorithms in PyTorch. We adopted the low-rank adaptation (LoRA) \cite{LoRA} for parameter efficient fine-tuning, applying LoRA parameters in parallel with all q and v parameters in the pre-trained Transformer \cite{MeSAM}, and set the low-rank hyperparameter $r$ to 4 as suggested in the original paper. Thus, our trainable parameters consist only of LoRA and the Head and Tail. We fine-tuned on the ARAD\_1K-Train dataset for 2000 epochs and on the RESISC45 dataset for 50 epochs, considering the number of training samples. We used the Adam optimizer with a learning rate of 0.001 and the batch size was 64.

When testing on unseen data, we set $R$ to 50\% of the number of channels $L$. For EigenSR-$\beta$, the number of iterations $N_{it}$ was set to 5, and the constant $\lambda$ was empirically set to 0.8 for SR $\times 2$, and 0.4 for SR $\times 4$ and $\times 8$, respectively.

\begin{figure*}[!t]
\centering
\subfloat[]{\includegraphics[scale=0.35]{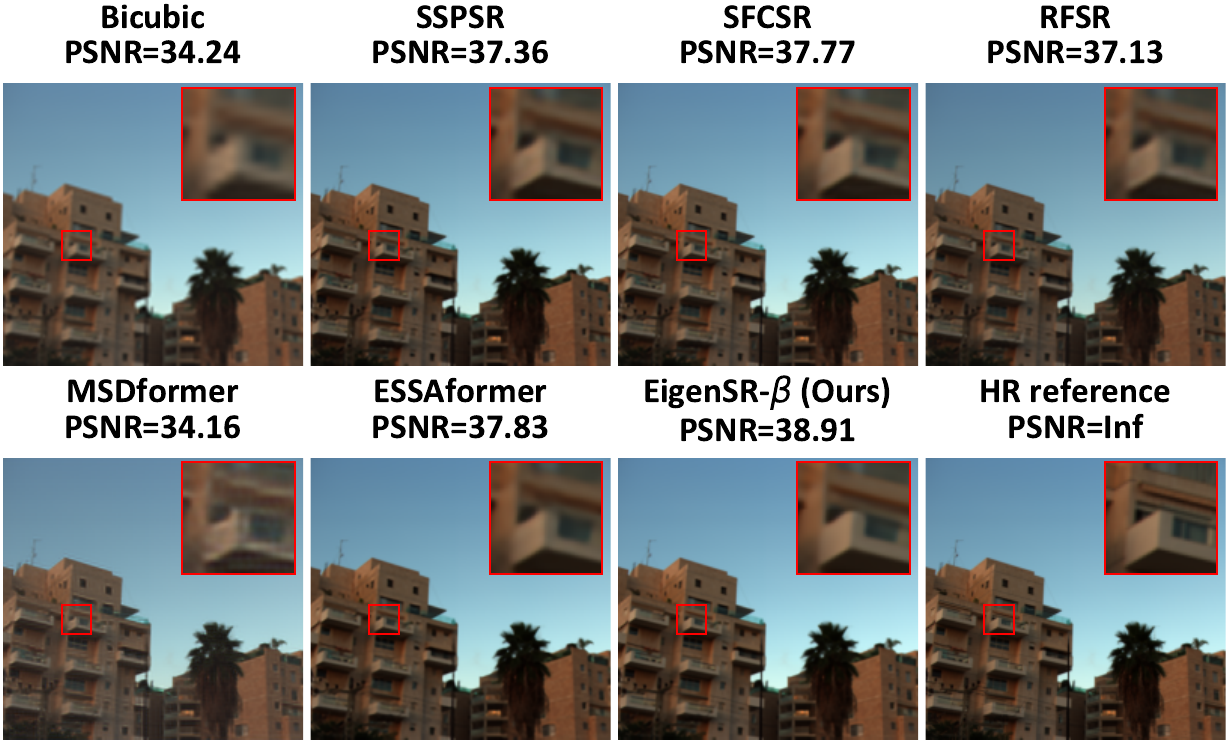}%
\label{fig_arad1k_sr}}
\hfil
\subfloat[]{\includegraphics[scale=0.35]{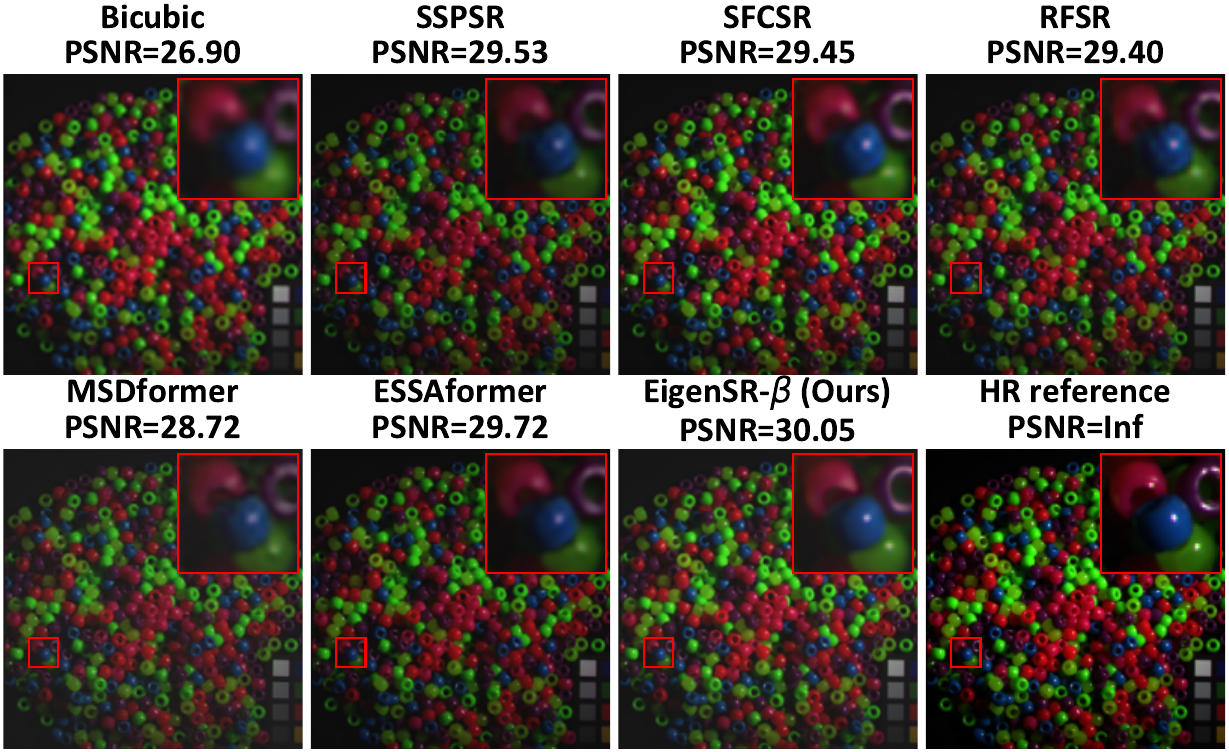}%
\label{fig_cave_sr}}
\vfil
\subfloat[]{\includegraphics[scale=0.35]{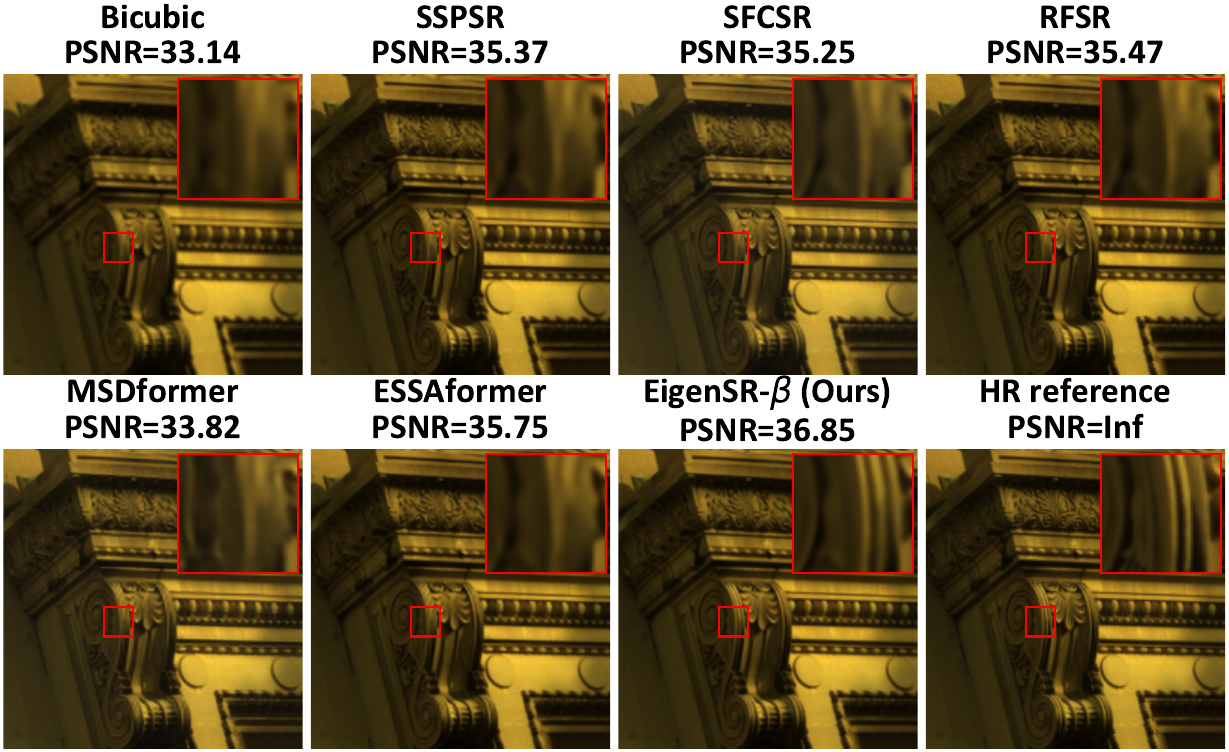}%
\label{fig_harvard_sr}}
\hfil
\subfloat[]{\includegraphics[scale=0.35]{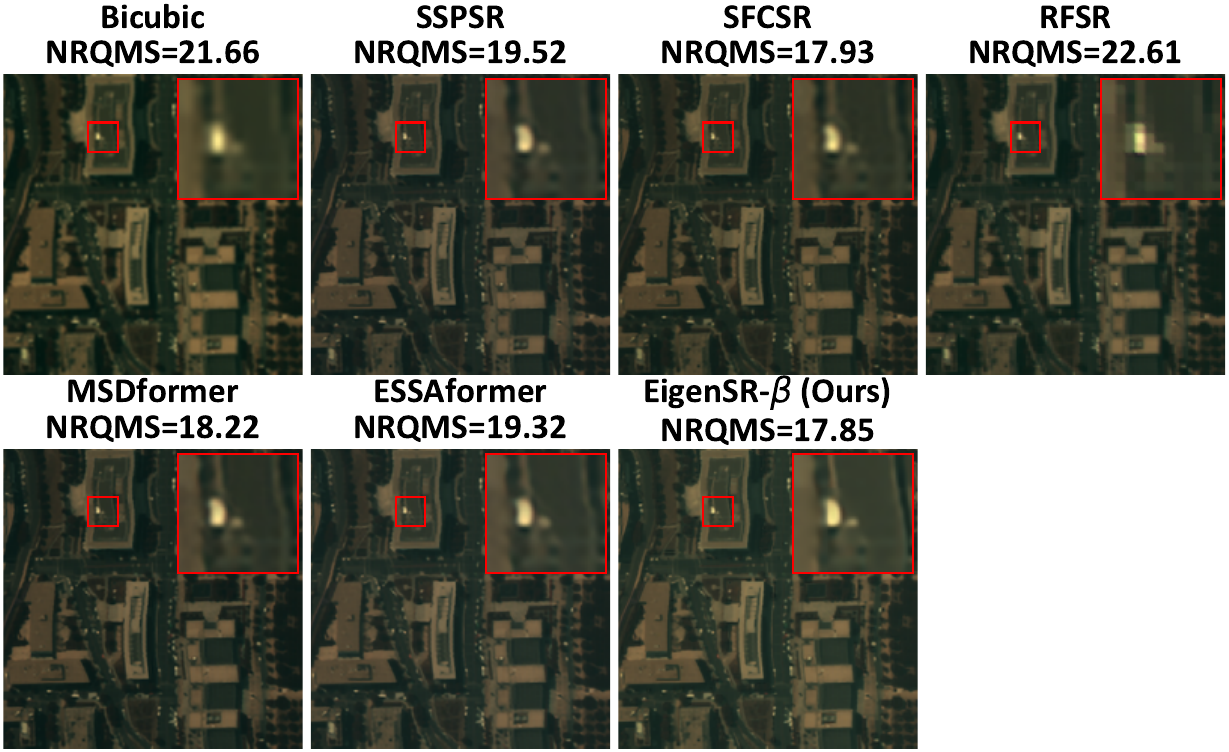}%
\label{fig_dcmall_sr}}
\caption{Visual results of SR $\times 4$. (a) ARAD\_1K-Test. (b) CAVE. (c) Harvard. (d) DC Mall. We used band numbers (27, 14, 6) to generate the pseudo-color images for ARAD\_1K, CAVE, and Harvard datasets, and band numbers (32, 14, 7) for DC Mall. } 
\label{fig:vis_comp}
\end{figure*}

\subsection{Comparisons with the SOTA Methods}

\subsubsection{Quantitative Comparisons.} First, we conducted comparisons for SR at $\times 2$, $\times 4$, and $\times 8$ scales with HR reference images. The LR images were obtained by downsampling the HR reference images. The supervised methods were trained on ARAD\_1K-Train, and all methods were tested on ARAD\_1K-Test, CAVE, and Harvard datasets. We used the PSNR, SSIM, and SAM to evaluate the SR quality. The results are shown in Table \ref{tab:sota_ref}. Our EigenSR-$\alpha$ and EigenSR-$\beta$ models achieved second and first rankings in both spatial and spectral metrics, respectively, in almost all cases. Since our models were fine-tuned on ARAD\_1K-Train, they also performed well on the CAVE and Harvard datasets, which were captured by different sensors, demonstrating their generalization ability to spectral-agnostic unseen data.

Second, we conducted blind SR tests at scale $\times 4$ without reference images on the LR remote sensing datasets Pavia, DC Mall, and Chikusei. We trained on the HR remote sensing RGB dataset RESISC45, due to the lack of HR remote sensing HSI for training. For those HSI-specific methods, we trained a 3-channel model and performed SR on the test HSI by processing adjacent three channels at a time. In contrast, since our single-channel model does not face channel alignment issues, we directly fine-tuned on the eigenimages of RGB images and tested on the eigenimages of HSI. We used the no-reference metric NRQMS to jointly evaluate the spatial and spectral quality, and recorded the inference time (in seconds), as shown in Table \ref{tab:sota_noref}. Our models can provide SR HSI with better spatial and spectral quality. Meanwhile, EigenSR-$\alpha$ consumes less inference time due to its efficiency in the eigenimage domain, while EigenSR-$\beta$ still keeps an acceptable inference time among other methods.

\begin{table}[t]
\renewcommand{\arraystretch}{0.6}
\scriptsize
\tabcolsep=0.05cm

% \resizebox{\columnwidth}{!}{
\begin{tabular}{l|c c|c c|c c}
\toprule
\multirow{2}{*}{\textbf{Method}} & \multicolumn{2}{c|}{\textbf{Pavia}} & \multicolumn{2}{c|}{\textbf{DC Mall}} & \multicolumn{2}{c}{\textbf{Chikusei}}\\
% \cline{2-7}
& \textbf{NRQMS}$\downarrow$ & \textbf{Time(s)} & \textbf{NRQMS}$\downarrow$ & \textbf{Time(s)} & \textbf{NRQMS}$\downarrow$ & \textbf{Time(s)} \\
\midrule

\rowcolor{gray!10}
Bicubic                            &22.42&-  &22.26&-  &24.70&- \\
DHSP \shortcite{DHSP}              &19.78&949.44 &18.93&724.10  &22.84&482.28 \\
\rowcolor{gray!10}
SSPSR \shortcite{SSPSR}            &19.55&\underline{5.24}  &19.34&\underline{13.48}  &21.57&\underline{9.83} \\
SFCSR \shortcite{SFCSR}            &\underline{19.19}&5.30  &18.39&23.76  &20.75&17.21 \\
\rowcolor{gray!10}
RFSR \shortcite{RFSR}              &22.90&\textbf{3.66}  &22.80&\underline{10.30}  &24.70&\underline{7.76} \\
TSBSR \shortcite{SR_Blind_SwinIR_SHISR}  &19.83&211.86 &19.43&159.28 &21.22&107.91 \\
\rowcolor{gray!10}
MSDformer \shortcite{MSDformer}    &19.33&8.88  &\underline{18.33}&28.52  &\underline{20.72}&20.31 \\
ESSAformer \shortcite{ESSAformer}  &20.04&21.34  &19.65&64.92  &22.12&44.50 \\
\rowcolor{gray!10}
DKP \shortcite{DKP_CVPR24}         &20.05&1205.31 &19.89&886.52 &22.98&450.40 \\
\textbf{EigenSR-}$\bm{\alpha}$ \textbf{(Ours)} &\underline{19.17}&\underline{4.79} &\underline{18.05}&\textbf{4.76} &\underline{20.58}&\textbf{3.36} \\
\rowcolor{gray!10}
\textbf{EigenSR-}$\bm{\beta}$  \textbf{(Ours)} &\textbf{19.13}&24.54  &\textbf{17.98}&24.36 &\textbf{20.42}&17.16 \\

\bottomrule
\end{tabular}
% }
\caption{Quantitative comparison without reference image at scale $\times 4$. RESISC45 is used for training. The best results are marked in \textbf{bold}, and the second and third are \underline{underlined}. }
\label{tab:sota_noref}
\end{table}

\subsubsection{Qualitative Comparisons.} We visually present several SR test results from a spatial view, in Figure \ref{fig:vis_comp}. Overall, our method can provide clearer and sharper SR results. For example, the stripes in the Harvard dataset and the small white target in the DC Mall dataset. This is because our method effectively introduces the model pre-trained on a large amount of RGB data, addressing the HSI data bottleneck, and enhancing spatial processing capabilities.

\begin{figure}[!t]
\centering
\includegraphics[scale=0.32]{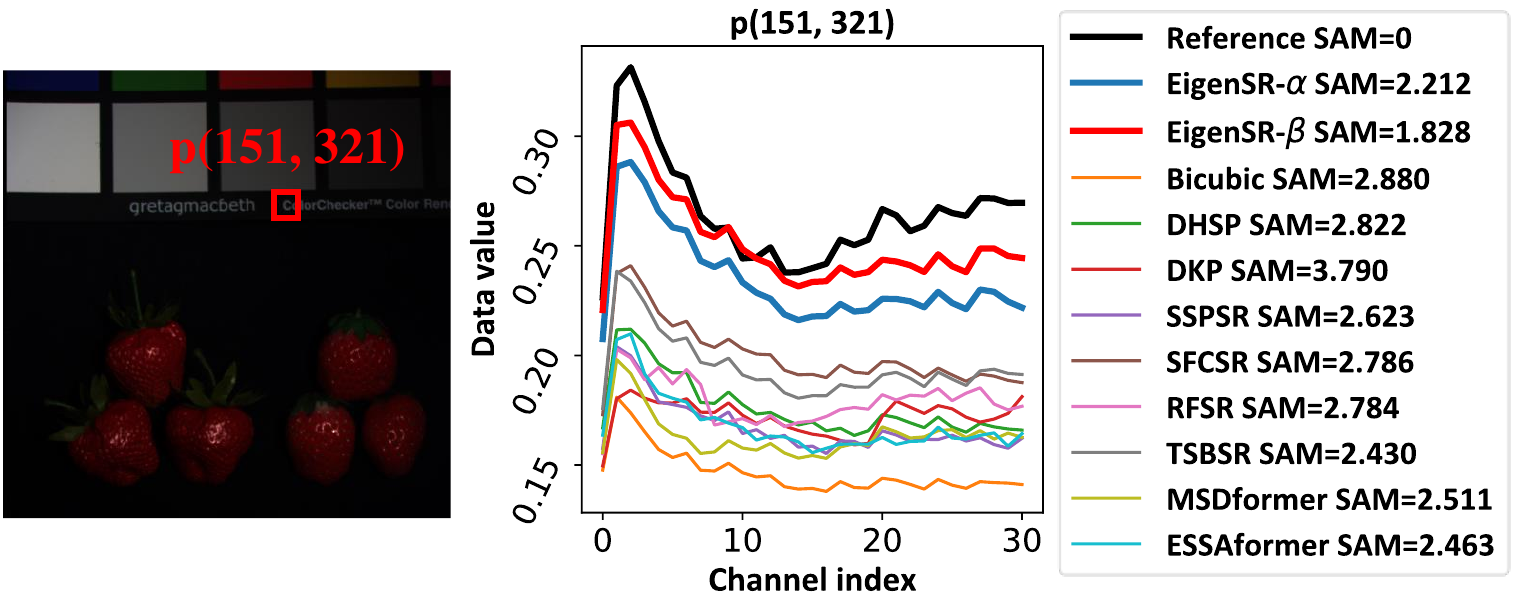}
\caption{SR $\times 4$ comparison in the spectral dimension on CAVE dataset. We denote the spatial coordinate as p(151, 321). Our method can provide more accurate spectral curves.}
\label{fig:cave_spec}
\end{figure}

On the other hand, we compared the spectra of SR HSI in Figure \ref{fig:cave_spec}. It can be seen that our method provides spectral curves closer to the reference spectral curve. This is because we preserved the spectral information during the SR of eigenimages and recovered it during reconstruction, and further constrained it through ISR. These resulted in more accurate spectral estimates.

\subsection{Ablation Studies} 

We conducted ablation studies on the ARAD\_1K dataset. The models were trained on ARAD\_1K-Train and tested on ARAD\_1K-Test at a scale factor of $\times 4$.

\begin{table}[t]
\centering
\renewcommand{\arraystretch}{0.6}
\scriptsize
% \small
\tabcolsep=0.06cm
% \resizebox{\columnwidth}{!}{
\begin{tabular}{l c c c c|c c c c}
\toprule
\textbf{Setting (explanation)} & \textbf{PT} & \textbf{NMF} & \textbf{Eigen} & \textbf{ISR} & \textbf{PSNR}$\uparrow$ & \textbf{SSIM}$\uparrow$ & \textbf{SAM}$\downarrow$ & \textbf{Time(s)} \\
\midrule
%                                   & PT        & NMF       & Eigen     & ISR       &
\rowcolor{gray!10}
\ding{172} (IPT baseline)           & \ding{51} &           &           &           &39.96&0.9556&1.268&3.19 \\
\ding{173} (IPT w/ NMF)             & \ding{51} & \ding{51} &           &           &39.81&0.9548&1.280&36.17 \\
\rowcolor{gray!10}
\ding{174} (EigenSR-$\alpha$)       & \ding{51} &           & \ding{51} &           &40.28&0.9602&1.223&\textbf{1.13} \\
\ding{175} (EigenSR-$\alpha$ w/o PT)&           &           & \ding{51} &           &39.74&0.9538&1.310&\textbf{1.13} \\
\rowcolor{gray!10}
\ding{176} (EigenSR-$\beta$)        & \ding{51} &           & \ding{51} & \ding{51} &\textbf{40.46}&\textbf{0.9605}&\textbf{1.182}&4.71 \\
\bottomrule
\end{tabular}
% }
\caption{Break-down ablation to improve the performance.}
\label{tab:full_ablation}
\end{table}

\begin{table}[!t]
\centering
\renewcommand{\arraystretch}{0.6}
\scriptsize
% \small
\tabcolsep=0.15cm
% \resizebox{\columnwidth}{!}{
\begin{tabular}{c|c c c c}
\toprule
\textbf{How to fine-tune} & \textbf{PSNR}$\uparrow$ & \textbf{SSIM}$\uparrow$ & \textbf{SAM}$\downarrow$ & \textbf{\#trainable params. (M)} \\
\midrule
Only Head \& Tail   &39.92&0.9551&1.250 & 0.78 \\
Head \& Tail + LoRA &40.46&0.9605&1.182 & 1.11 \\
Full parameter      &40.60&0.9610&1.175 & 115.68 \\
\bottomrule
\end{tabular}
% }
\caption{Different fine-tuning architectures.}
\label{tab:finetune}
\end{table}

\subsubsection{Break-down Ablation.} We validated each component of our method by breaking them down, as shown in Table \ref{tab:full_ablation}. We set the pre-trained (PT) model IPT followed by fine-tuning on HSIs, as the baseline (\ding{172}). The additional NMF spectral constraint \cite{SR_Blind_SwinIR_SHISR} was applied during inference (\ding{173}). Differently, EigenSR-$\alpha$ (\ding{174}) fine-tuned the pre-trained model in the eigenimage domain (Eigen). We removed the pre-trained weights (\ding{175}) from EigenSR-$\alpha$ and trained the model from scratch instead of fine-tuning. The ISR was further added to EigenSR-$\alpha$ as the EigenSR-$\beta$ (\ding{176}).

It can be found that NMF leads to a performance decrease compared to the baseline while incurring a non-negligible time cost during inference. This is probably because the NMF optimization has the non-convex nature, making it difficult to adapt to various situations. In contrast, EigenSR-$\alpha$ outperforms both \ding{172} and \ding{173} while maintaining efficient inference, which validates its effectiveness of learning the eigenimages. Leveraging the RGB pre-trained weights results in an improvement of \ding{174} over \ding{175}. Moreover, despite introducing a longer but still acceptable inference time, EigenSR-$\beta$ further outperforms EigenSR-$\alpha$ in terms of performance.

\subsubsection{Fine-Tuning Architectures.} We studied three different fine-tuning architectures in Table \ref{tab:finetune}: only fine-tuning the Head \& Tail, fine-tuning the Head \& Tail along with the LoRA embedded in the Transformer, and full parameter fine-tuning. It can be observed that full parameter fine-tuning leads to the best performance but imposes a heavy burden in terms of parameters. Only using the Head \& Tail for fine-tuning does not achieve optimal performance. Incorporating LoRA allows the Transformer to better adapt to the eigenimage domain while keeping parameter efficiency.

\begin{figure}[!t]
  \centering
  % \hfill
  \subfloat[]{\includegraphics[scale=0.33]{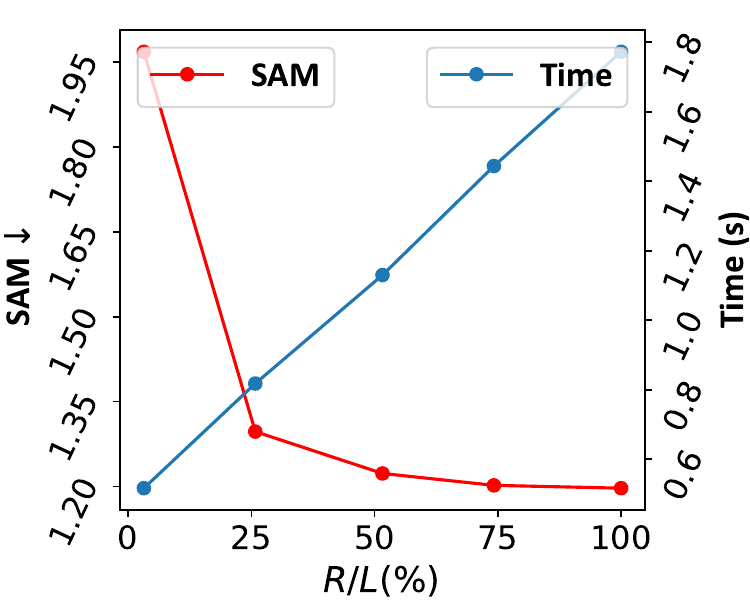}%
  \label{fig_R}}
  \subfloat[]{\includegraphics[scale=0.33]{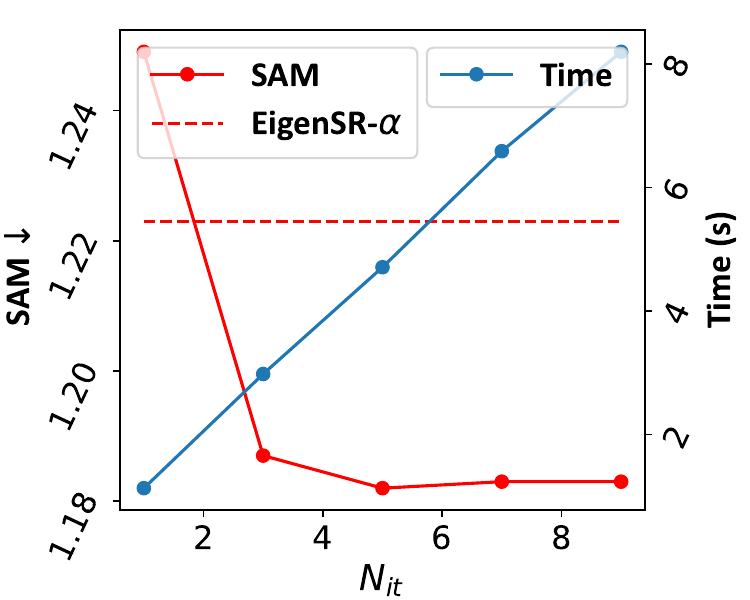}%
  \label{fig_Nit}}
  \caption{Effects of inference hyperparameters. (a) Rank $R$. (b) Iteration number $N_{it}$ ($\lambda = 0.4$ for EigenSR-$\beta$). }
  \label{fig:hyperparam}
\end{figure}

\subsubsection{Inference Hyperparameters.} The rank $R$ and the iteration number $N_{it}$ balance the SR performance and time costs. Figure \ref{fig:hyperparam} studies the effects of the two hyperparameters. It can be observed that as the two hyperparameters increase, the performance gradually improves and eventually saturates, while the inference time grows linearly. Considering the balance between performance and time, we suggest a reasonable configuration: $R/L = 50\%$ and $N_{it} = 5$.

\section{Conclusion}

This work presents EigenSR for single-HSI-SR, a method that fully leverages the spatial capabilities of pre-trained RGB models while maintaining spectral correlations, to address the data bottleneck of HSI. It is based on the concept of spatial-spectral decoupling via SVD and consists of two stages. First, we fine-tune a single-channel model on the spatial components, i.e., the eigenimages, allowing the RGB model channels to align with HSI and adapt to the eigenimage domain. Then, during inference on unseen data, we decompose the LR HSI, perform channel-wise SR on its eigenimages, and reconstruct the SR HSI using the spectral components, i.e., the spectral basis, to maintain spectral correlations. Furthermore, we propose ISR to fully exploit the LR HSI and enhance spectral fidelity. We only introduce the pre-trained model in the spectral-agnostic spatial components, which enhances the model's generalization ability. Additionally, by utilizing a small number of low-rank eigenimages, we reduce the inference complexity. Extensive experiments demonstrate that our method outperforms SOTA approaches, proving the effectiveness of introducing pre-trained RGB models to address the data scarcity in HSI.

\section{Acknowledgments}
This work was supported by the National Natural Science Foundation of China (U2133211, 62301092, 62301093). Xi~Su expresses special gratitude to Dr. Shuanghao~Bai for providing valuable feedback on the writing of this paper.

% \bibliography{aaai25}

\newpage
\section{Appendix}

\renewcommand{\thefigure}{a\arabic{figure}}
\renewcommand{\thetable}{a\arabic{table}}
\renewcommand{\theequation}{a\arabic{equation}}
\setcounter{figure}{0}
\setcounter{table}{0}
\setcounter{equation}{0}

\subsection{Proof of Theorem 1}

\noindent\textbf{Theorem 1}: For eigenimages $\mathbf{E} = \mathbf{U}^{\top} \mathbf{Y}$, we have:	
\begin{enumerate}
\item If two column vectors $\mathbf{Y}_{:, i}$ and $\mathbf{Y}_{:, j}$ of $\mathbf{Y}$ are identical, then their corresponding $\mathbf{E}_{:, i}$ and $\mathbf{E}_{:, j}$ are also identical.
\item If $\mathbf{Y}$ is downsampled by a linear operator $\mathbf{D}$, then the same $\mathbf{D}$ concurrently acts on $\mathbf{E}$.
\end{enumerate}	 

The following contents provide the proof of Theorem 1.

\begin{enumerate}
\item \textbf{Proof}: Suppose that $\mathbf{Y}_{:, i}$ and $\mathbf{Y}_{:, j}$ are two identical vectors and have their corresponding $\mathbf{E}_{:, i}$ and $\mathbf{E}_{:, j}$ under $\mathbf{U}$, i.e., the following equation holds:
\begin{equation}
\mathbf{0} = \mathbf{Y}_{:, i} - \mathbf{Y}_{:, j} = \mathbf{U} \mathbf{E}_{:, i} - \mathbf{U} \mathbf{E}_{:, j} = \mathbf{U} (\mathbf{E}_{:, i} - \mathbf{E}_{:, j}).
\end{equation}
Since $\mathbf{U}$ is a semiunitary matrix satisfying $\mathbf{U}^{\top} \mathbf{U} = \mathbf{I}$, we obtain the following by left-multiplying both sides of the equation by $\mathbf{U}^{\top}$:
\begin{equation}
\mathbf{0} = \mathbf{U}^{\top} \mathbf{U} (\mathbf{E}_{:, i} - \mathbf{E}_{:, j}) = \mathbf{E}_{:, i} - \mathbf{E}_{:, j}.
\end{equation}
Thus we derive $\mathbf{E}_{:, i} = \mathbf{E}_{:, j}$.

\item  \textbf{Proof}: Suppose that the degradation operator $\mathbf{D}\in\mathbb{R}^{N\times \left (N/\kappa^{2}  \right )}$ holds in original space $\mathbf{Y}$, i.e., $\mathbf{Y}_{LR} = \mathbf{Y}_{HR} \mathbf{D}$. Then, it holds in the eigenimage space $\mathbf{E}$: 
\begin{equation}
\begin{aligned}
\mathbf{E}_{LR} &= \mathbf{U}^{\top} \mathbf{Y}_{LR} = \mathbf{U}^{\top} (\mathbf{Y}_{HR} \mathbf{D}) =  (\mathbf{U}^{\top} \mathbf{Y}_{HR} )\mathbf{D} \\
&= \mathbf{E}_{HR} \mathbf{D}.
\end{aligned}
\end{equation}
\end{enumerate}

These complete the proof.

\subsection{How to Select a Channel in the Fine-Tuning Stage}

\begin{figure}[!t]
  \centering
  \includegraphics[width=\columnwidth]{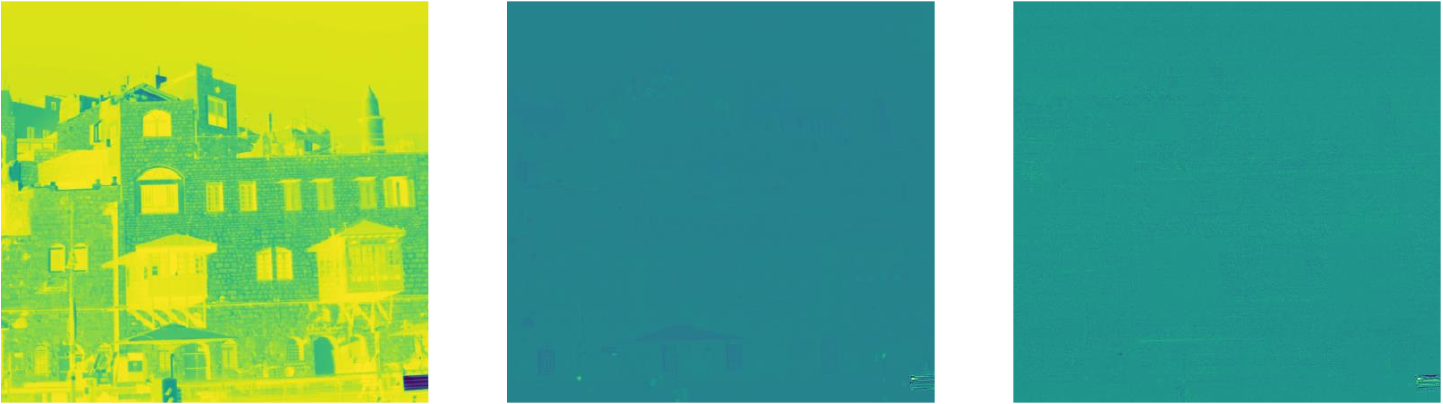}
    \caption{Eigenimages of the ARAD\_1K-0786 image. The channel indices from left to right are 1, 16, and 31, respectively.}
  \label{fig:arad1k_eigenimages}
\end{figure}

\begin{figure}[!t]
  \centering
  \includegraphics[width=\columnwidth]{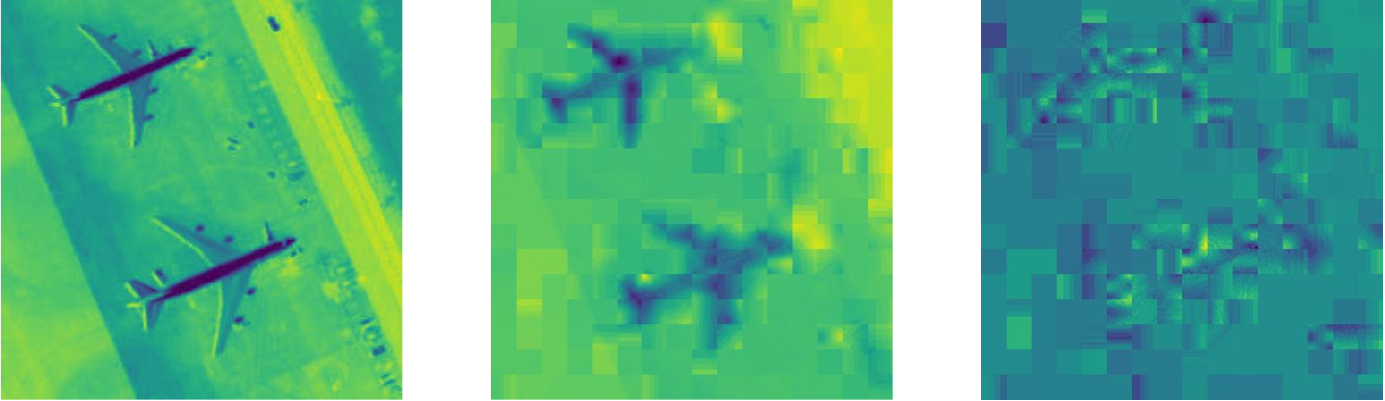}
    \caption{Eigenimages of the RESISC45-airplane012 image. The channel indices from left to right are 1, 2, and 3, respectively.}
  \label{fig:resisc45_eigenimages}
\end{figure}

We found that randomly selecting a channel $c$ from all channels of the eigenimages in \textbf{\textit{Step 2}} of the fine-tuning stage does not achieve optimal performance. As shown in Figures \ref{fig:arad1k_eigenimages} and \ref{fig:resisc45_eigenimages}, channels with higher numbers exhibit lower image signal-to-noise ratios or more compression artifacts, which are detrimental to training that requires high-quality images. This is because the energy of each eigenimage channel for reconstructing the original image decreases as the channel number increases, which is reflected by the singular values $\sigma$ arranged in descending order, as shown in Eq. (2). As a result, the quality of the eigenimages decreases as the channel number increases.

Therefore, we control the quality of the eigenimages used for training by setting a threshold related to the magnitude of $\sigma$. However, assigning fixed $\sigma$ thresholds to all HSIs is unfeasible on the whole datasets as each HSI may contain eigenimages of varying importance. To address this, we establish a threshold value $\tau$ that regulates the cumulative energy of $\sigma$ at the training set level, instead of the image level. Then, eigenimage channels below this threshold are then utilized for training. Specifically, a cutoff channel $p$ is given as the following:
\begin{equation}
\label{eq:threshold}
p = \underset{\substack{j \in \{1, \ldots, L\}, \\ q_j \leq \tau}}{{\arg\max} \, } q_j = \text{cumsum}(\frac{\sigma}{\sum_i^L \sigma_i})_j,
\end{equation}
where $\text{cumsum}(\cdot)$ calculates the cumulative sum of a vector. After obtaining $p$ for one HSI, we randomly choose one channel $c$ from $\text{Uniform}(\{1, \ldots, p\})$ to feed the single-channel model for fine-tuning. 

\begin{figure}[!t]
  \centering
  \subfloat[]{\includegraphics[scale=0.33]{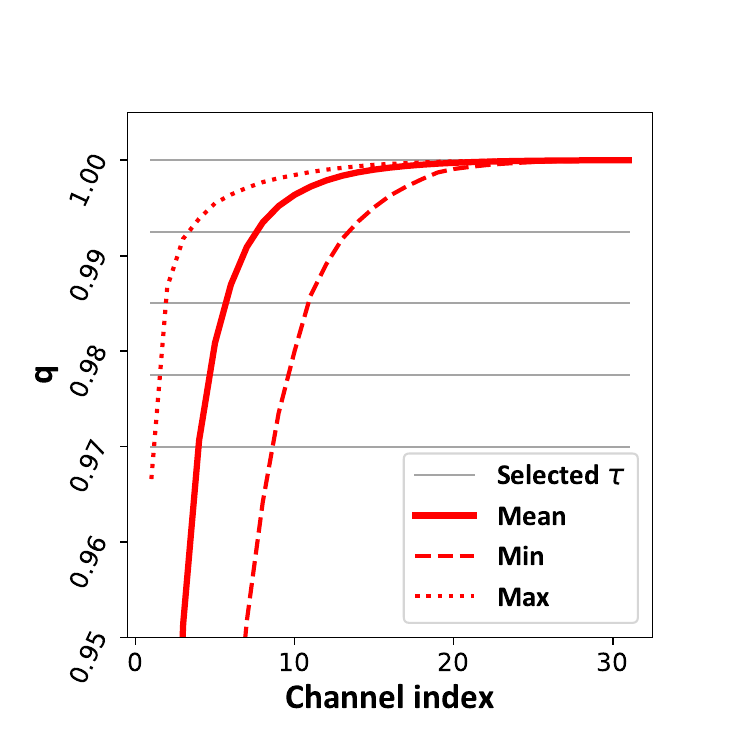}%
  \label{fig_q_tau}}
  % \hfill
  \subfloat[]{\includegraphics[scale=0.33]{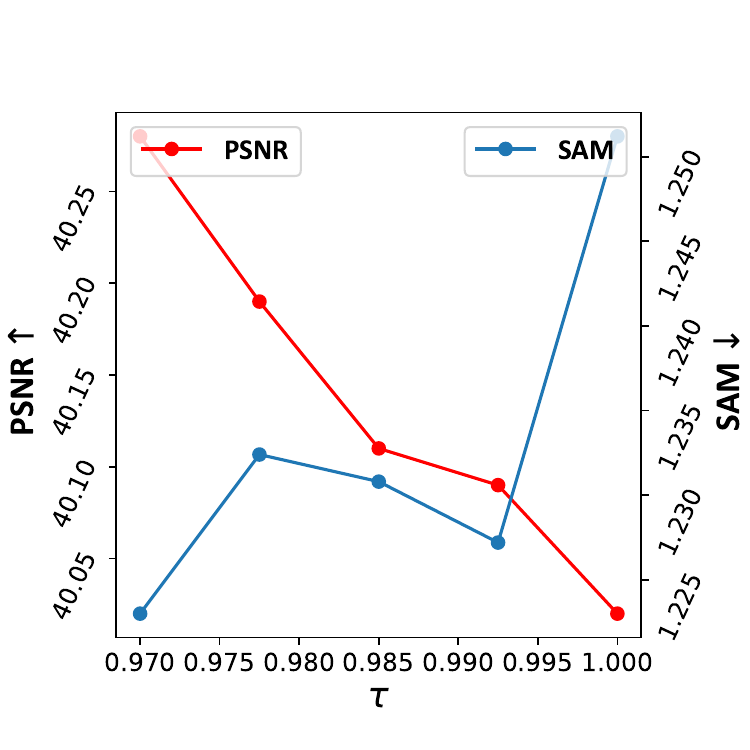}%
  \label{fig_arad1k_tau}}
  \caption{Selecting the channels of training images during the fine-tuning stage. (a) The cumulative curves of singular values $\sigma$ on the ARAD\_1K-Train and five tested threshold values $\tau$. (b) PSNR and SAM with respect to $\tau$. A lower $\tau$ (main components at the dataset level) provides better performance.}
  \label{fig:fine-tuning_rank}
\end{figure}

The cumulative curves of $q_j$ on the entire training set of the ARAD\_1K are shown in Figure \ref{fig:fine-tuning_rank}(a). We tested five $\tau$ values ranging from 0.97 to 1.00 in Figure \ref{fig:fine-tuning_rank}(b) and observed that a lower $\tau$ leads to improved performance. This means that retaining only a small number of main components at the training set level during fine-tuning is more useful for learning. Note that the selection of $\tau$ is constrained to ensure that the number of channels in all training set eigenimages remains at least 1, as depicted by the Max curve in Figure \ref{fig:fine-tuning_rank}(a). Currently, the $\tau$ value is around 0.97.

For fine-tuning on the RGB dataset RESISC45, we used only the first eigenimage channel, as the eigenimages of these images contain only three channels, and the quality of the latter two channels is low.

\end{document}